\documentclass[
 superscriptaddress,
 amsmath,
 amssymb,
 aps,
 a4paper,
 floatfix,
 longbibliography,
 pra,
 twocolumn
]{revtex4-2}

\usepackage{graphicx}
\usepackage{dcolumn}
\usepackage{bm}

\usepackage{physics}

\usepackage{xcolor}

\usepackage{microtype}

\usepackage[american]{babel}

\usepackage[utf8]{inputenc}
\usepackage{textcomp}
\usepackage{siunitx} 
\usepackage{booktabs}
\usepackage{multirow}

\DeclareUnicodeCharacter{2212}{\textendash}
\newcommand{\ee}{\text{e}}
\newcommand{\ii}{\text{i}}
\newcommand{\tauprime}{\tau^{\prime}}

\newcommand{\tauzero}{\tau_{0}}

\newcommand{\limtauzeroinfty}{\lim_{\tauzero\rightarrow\infty}}

\newcommand{\tzero}{t_{0}}
\newcommand{\tildea}{\tilde{a}}

\newcommand{\orcid}[1]{\href{https://orcid.org/#1}{\includegraphics[width=7pt]{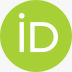}}}

\definecolor{cset-aps-blueberry}{RGB}{28,128,158}
\definecolor{cset-aps-blue}{RGB}{46,44,184}
\definecolor{cset-aps-turquoise}{RGB}{0,67,88}
\definecolor{cset-aps-limegreen}{RGB}{190,219,67}
\definecolor{cset-aps-green}{RGB}{31,138,112}
\definecolor{cset-aps-yellow}{RGB}{255,225,25}
\definecolor{cset-aps-orange}{RGB}{253,116,0}
\definecolor{cset-aps-red}{RGB}{219,0,43}
\definecolor{myred}{RGB}{255,0,20}

\usepackage{hyperref}
\hypersetup{%
    colorlinks=true,
    linkcolor={cset-aps-red},
    linkbordercolor={cset-aps-red},
    filecolor={cset-aps-orange},
    filebordercolor={cset-aps-orange},
    citecolor={cset-aps-blue},
    citebordercolor={cset-aps-blue},
    urlcolor={cset-aps-blueberry},
    urlbordercolor={cset-aps-blueberry},
    menucolor={cset-aps-limegreen},
    menubordercolor={cset-aps-limegreen},
    breaklinks=true,
    pdfborderstyle={/S/U/W 2},
    pdfpagemode=UseOutlines,
    pdfstartpage={1},
}

\usepackage[capitalise]{cleveref}

\AtBeginDocument{\RenewCommandCopy\qty\SI}

\begin{document}

\preprint{APS/123-QED}

\title{A logarithmic phase singularity at the heart of Landau-Zener transitions}

\author{Eric P. Glasbrenner \orcid{0000-0002-0822-3888}}
\affiliation{Institut f{\"u}r Quantenphysik and Center for Integrated Quantum
    Science and Technology (IQST), Universit{\"a}t Ulm, Albert-Einstein-Allee 11, D-89081 Ulm, Germany}
\email{eric.glasbrenner@uni-ulm.de}
\author{David Fabian}%
\affiliation{Institut f{\"u}r Quantenphysik and Center for Integrated Quantum
   Science and Technology (IQST), Universit{\"a}t Ulm, Albert-Einstein-Allee 11, D-89081 Ulm, Germany}
\email{dfabian@mailbox.org}
\author{Wolfgang P. Schleich \orcid{0000-0002-9693-8882}}
\affiliation{Institut f{\"u}r Quantenphysik and Center for Integrated Quantum
    Science and Technology (IQST), Universit{\"a}t Ulm, Albert-Einstein-Allee 11, D-89081 Ulm, Germany}
\affiliation{Institute for Quantum Science and Engineering (IQSE), and Texas A\&M AgriLife Research and Hagler Institute for Advanced Study, Texas A\&M University, College Station, TX 77843-4242, USA}

\date{\today}
\begin{abstract}
Three ingredients of the elementary Landau-Zener problem determine the familiar expression $a_{LZ}\equiv\exp\left[-\pi/(2\epsilon)\right]$ for the asymptotic value of the probability amplitude for remaining in the initial level:
(i) A wave whose phase is determined by the product of a contour integral over a simple pole at the origin of the complex plane and the inverse of twice the scaled chirp parameter $\epsilon$. 
(ii) An asymptotic limit of the associated path connecting the points $\pm 1$ along the real axis and circumventing the pole in the upper half-plane, and (iii) a half-circle in the lower half plane enclosing together with the asymptotic path the pole.
The Cauchy theorem immediately provides us with the value $\ii\pi$ of the asymptotic contour, and thus with $a_{LZ}$. Our analysis demonstrates not only that $a_{LZ}$ is the consequence of a logarithmic phase singularity but also explains why the Markov approximation also leads to $a_{LZ}$. 
\end{abstract}


\maketitle

\section{Introduction}
\label{section:introduction}
Logarithmic phase singularities occur in diverse areas of physics and can lead to subtle quantum phenomena. 
For example, the Hawking-Unruh radiation \cite{Hawking1974, Unruh1976, Hawking1977} is a consequence of such a singularity \cite{scully2018} appearing in the mode functions of the electromagnetic field at the event horizon.
Moreover, the energy eigenfunctions of an inverted harmonic oscillator \cite{Ullinger2022} display a logarithmic phase singularity in phase space promoting the separatrices to event horizons \cite{Rozenman2024}.
In this article, we identify a logarithmic phase singularity in the Landau-Zener problem, and use it to rederive the familiar probability amplitude \cite{Landau1932b, Zener1932, Majorana1932, Stueckelberg1932}
\begin{align}
    \label{intro:eq:LZ}
    a_{LZ} \equiv \exp\left(-\frac{\pi}{2\epsilon}\right)
\end{align}
where $\epsilon$ is the scaled chirp parameter. 
\par
The Landau-Zener framework plays a central role in non-adiabatic transition physics, and despite its idealizations and many extensions \cite{Dykhne1960, Suominen1991, Garraway1992, Akulin1992, Vitanov1999, Liu2002, nakamura2012, Shytov2004}, it remains widely applicable \cite{Peik1997, Ankerhold2003, Saito2007, Burrows2017, Gebbe2021, Fitzek2024, kofman2024, Konrad2024, bjoerkman2025, shendryk2025, lima2025} and is a standard benchmark for approximation methods \cite{Berry1993, Kayanuma1997, Kofman2023, Guttieres2023, Sun2025}.
Numerous techniques have been employed to redrive \cref{intro:eq:LZ} and range from asymptotic expansions of the exact solution in terms of parabolic cylinder functions \cite{Vitanov1996, SHEVCHENKO2010, Torosov2017, Glasbrenner2025, Glasbrenner2026}, via paths in the complex plane and the Markov approximation \cite{Glasbrenner2023, Glasbrenner2024book}, to approximation schemes based on the Wentzel-Kramers-Brillouin wave functions \cite{Dykhne1962, Davis1976, Teranishi1997, Enomoto2022} and the change of the behavior of a function when crossing a Stokes lines, just to name a few.
For a more comprehensive treatment, see reference \cite{ivakhnenko2023}.
However, to the best of our knowledge, the important role of the asymptotic time in determining \cref{intro:eq:LZ} has not been appreciated. 

Indeed, the standard approach towards the Landau-Zener problem starts from the quantum dynamics of the two probability amplitudes between a \textit{finite} initial time and a \textit{finite} final time. 
The familiar expression, \cref{intro:eq:LZ}, for $a_{LZ}$ only emerges when these times tend towards infinity, thus covering the dynamics from the infinite past to the infinite future. 

In the present article, we perform first this limit, and then consider the resulting quantum dynamics. 
This interchange of the dynamics and the limit not only simplifies the underlying equations considerably but brings to light the crucial role of the logarithmic phase singularity in determining the Landau-Zener expression.
Although this interchange is \textit{not} justified by the equations it still produces the correct answer.

In order to verify this claim, we express the dynamics in terms of a time-dependent path in the complex plane with a simple pole at the origin providing us with the logarithmic phase. 
This representation allows us to analyze the dependence of the path on the initial and the final time.
When these times approach infinity, we arrive indeed at the equations resulting from the interchange of the dynamics and the limit.

Our article is organized as follows:
In section \ref{section:2}, we first formulate the Landau-Zener problem, scale the underlying equations, and then perform the limit as to cover the dynamics from the infinite past to the infinite future. 
This analysis leads us immediately to a logarithmic phase.

We dedicate section \ref{section:3} to a generalization of this ansatz and define the probability amplitude in terms of a contour integral. 
The corresponding path obeys a nonlinear differential equation of second order but approaches an elementary path in the asymptotic limit as demonstrated by numerical as well as approximate analytical solutions.
When we apply the Cauchy theorem to the asymptotic path, we recover immediately the Landau-Zener result, \cref{intro:eq:LZ}.

In section \ref{section:4}, we answer the question \cite{Glasbrenner2023}: Why is the Markov approximation exact?
We conclude in section \ref{section:5} by summarizing our results and providing an outlook.
Finally, in appendix \ref{appendix:A}, we derive approximate analytical expressions for the asymptotic approach of the end points of the exact and the Markov paths. 

\section{Logarithmic phase ansatz}
\label{section:2}

In mathematics it is well-known that interchanging a differentiation and a limit is a dangerous enterprise.
However, in physics it is often considered a mathematical subtlety, which can be ignored since it is hardly ever of importance.

In the present section, we show that such an interchange in the Schrödinger equations of the Landau-Zener problem provides us with an immediate insight and identifies a logarithmic phase singularity as the origin of the Landau-Zener formula, \cref{intro:eq:LZ}.
At the same time we express a stern warning against this procedure and use the remainder of the article to verify the conclusions obtained in this heuristic way.

\subsection{First dynamics and then asymptotics:\\
An appropriate scaling}

In this section, we motivate our logarithmic phase ansatz for the Landau-Zener problem. 
For this purpose, we first formulate the problem, present a new scaling of the underlying equations and then transform them into a differential equation of second order. 

\subsubsection{Formulation of the problem}

The Landau-Zener problem involves in its most elementary version the \textit{linear} crossing of two energy levels with
a time-independent coupling. The corresponding Schrödinger equations for the two probability amplitudes $ a = a(t)$ and $b = b(t)$ read
\begin{align}
    \label{eq:a}
    \ii \dv{t} a &= -\alpha t a + \Omega b,\\
    \label{eq:b}
    \ii \dv{t} b &= \alpha t b + \Omega a.   
\end{align}

Here, $t$, $\alpha$ and $\Omega$ denotes time, the chirp involving the steepness of the level crossing and the coupling constant between the two levels, respectively.

Throughout the article, we consider the initial condition
\begin{align}
        \label{eq:initial:condition:a}
        a(-\tzero) = 1
\end{align}
which due to the conservation of probability
\begin{equation}
    \label{eq:conservation:prob}
    \abs{a}^2 + \abs{b}^2 = 1
\end{equation}
following from the unitary time evolution of the Schrödinger equations, \cref{eq:a,eq:b} enforces the identity 
\begin{align}
    \label{eq:initial:condition:b}
    b(-\tzero) = 0.
\end{align}

We solve \cref{eq:a,eq:b} subjected to the initial conditions, \cref{eq:initial:condition:a,eq:initial:condition:b}, for the finite time interval \mbox{$-\tzero\leq t \leq \tzero$}.
Hence, the solutions for the two probability amplitudes $a$ and $b$ depend on the choice of the initial time parameter $\tzero$ and the variable $t$. In order to bring out this fact, we include $t$ as well as $\tzero$ in the argument of \mbox{$a = a(t; \tzero)$} and \mbox{$b = b(t; \tzero)$}.

For a fixed value of $\tzero$, we determine $a$ from \cref{eq:a,eq:b} at $t=\tzero$, that is $a(\tzero, \tzero)$, and the Landau-Zener result, \cref{intro:eq:LZ}, only emerges in the limit $\tzero\rightarrow\infty$, that is 
\begin{align}
\label{eq:Lz:border}
    \lim_{\tzero\rightarrow\infty} a(\tzero; \tzero) = \exp\left(-\frac{\pi}{2\epsilon}\right) = a_{LZ},
\end{align}
where $\epsilon \equiv \alpha/\Omega^{2}$.

\subsubsection{Asymptotic scaling}

Therefore, it is useful to introduce the dimensionless time 
\begin{align}
    \label{eq:scaling:t:tzero}
    \tau \equiv \frac{t}{-\tzero}
\end{align}
which maps the time interval \mbox{$-\tzero\leq t \leq \tzero$} onto \mbox{$1 \geq \tau \geq -1$}. 
Obviously, the initial and final times $-\tzero$ and $\tzero$ correspond to $\tau = 1$ and $\tau = -1$.

Thus, \cref{eq:a,eq:b} take the form
\begin{align}
    \label{eq:a:t:zero}
    \ii \dot{a} &= -\epsilon\tauzero^{2}\tau a - \tauzero b\\
    \label{eq:b:t:zero}
    \ii \dot{b} &= \epsilon\tauzero^{2}\tau b - \tauzero a   
\end{align}
for the two probability amplitudes \mbox{$a = a(\tau; \tauzero)$} and \mbox{$b = b(\tau; \tauzero)$}.
Here, we have introduced the abbreviation $\tauzero \equiv \Omega\tzero$ and the dot denotes the derivative with respect to $\tau$.

Moreover, the initial conditions, \cref{eq:initial:condition:a,eq:initial:condition:b} now read
\begin{align}
    \label{eq:initial:condition:scaled:a}
    a(1; \tauzero) = 1
\end{align}
and 
\begin{align}
    \label{eq:initial:condition:scaled:b}
    b(1; \tauzero) = 0.
\end{align}

\subsubsection{Interaction picture}

In order to solve \cref{eq:a:t:zero,eq:b:t:zero}, we first integrate the terms proportional to $\tau$ and then consider the remaining dynamics due to the coupling. 
This procedure corresponds to a transformation into the interaction picture.

For this purpose, we now make the ansatz
\begin{align}   
    \label{eq:ansatz:tau:nu}
    a(\tau; \tauzero) \equiv \ee^{\ii\epsilon\tauzero^{2}(\tau^{2}-1)/2}\tildea(\tau; \tauzero)
\end{align}
with the initial condition 
\begin{align}
    \label{eq:initial:conditon:tilde:a:scaled}
    \tildea(1; \tauzero) = 1
\end{align}
following from \cref{eq:initial:condition:scaled:a,eq:ansatz:tau:nu}.

The relation
\begin{align}
    \label{eq:derivative:of:ansatz:tau:nu}
    \ii\dot{a} = -\epsilon\tauzero^{2}\tau a + \ee^{\ii\epsilon\tauzero^{2}(\tau^{2}-1)/2} \dot{\tildea},
\end{align}
and a comparison to \cref{eq:a:t:zero} leads us to the identification
\begin{align}
    \label{eq:b:atilde:dot}
    b(\tau; \tauzero) = -\frac{\ii}{\tauzero}\ee^{\ii\epsilon\tauzero^{2}(\tau^{2}-1)/2} \dot{\tildea}(\tau, \tauzero),
\end{align}
and with one more differentiation we arrive at the identity 
\begin{align}
    \label{eq:dot:b:tilde:ddot:a}
    \ii\dot{b} = -\epsilon\tauzero^{2}\tau b + \frac{1}{\tauzero} \ee^{\ii\epsilon\tauzero^{2}(\tau^{2}-1)/2} \ddot{\tildea}
\end{align}
where we have made use of \cref{eq:derivative:of:ansatz:tau:nu}.

Finally, a comparison to \cref{eq:b:t:zero} provides us with the differential equation 
\begin{align}
    \label{eq:differential:equation:a:tilde}
    \frac{1}{\tauzero^{2}}\ddot{\tildea} +2\ii\epsilon\tau\dot{\tildea} + \tildea = 0
\end{align}
of second order in time scaled by the initial time $\tauzero$.

Moreover, \cref{eq:initial:condition:scaled:b,eq:b:atilde:dot} yield the initial condition
\begin{align}
    \label{eq:initial:condition:dot:tilde:a}
    \dot{\tildea}(1; \tauzero) = 0.
\end{align}
Hence, \cref{eq:differential:equation:a:tilde} together with the initial conditions \cref{eq:initial:conditon:tilde:a:scaled} and \cref{eq:initial:condition:dot:tilde:a} determine the dynamics of $\tildea$.

\subsection{First asymptotics and then dynamics:\\ A first glimpse of the logarithm}

It is tempting to perform the asymptotic limit $\tauzero\rightarrow\infty$ in the differential equation, \cref{eq:differential:equation:a:tilde}, and to neglect the second derivative $\ddot{\tildea}$.
In this way, we arrive at the differential equation
\begin{align}
    \label{eq:differential:equation:a:tilde:limit}
    \tau\dot{\tildea}_a = \frac{\ii}{2\epsilon}\tildea_a
\end{align}
of first order in time for the asymptotic function
\begin{align}
    \tildea_a(\tau) \equiv \lim_{\tauzero\rightarrow\infty}\tildea(\tau; \tauzero).
\end{align}

We emphasize that, this approach assumes that we can interchange the differentiation with the limit. 
In the next section, we demonstrate that this often practiced trick is not valid in the present context due to the Stueckelberg oscillations \cite{Glasbrenner2025} appearing for negative values of $\tau$. 

However, \cref{eq:b:atilde:dot,eq:dot:b:tilde:ddot:a} show already why this interchange is problematic.
Due to the normalization condition, \cref{eq:conservation:prob}, we find $\abs{b}\leq 1$ for all values of $\tau$.
Thus, \cref{eq:b:atilde:dot} predicts $\abs{\dot{\tildea}}\sim\tauzero$, and from \cref{eq:dot:b:tilde:ddot:a} we find $\abs{\ddot{\tildea}}\sim\tauzero^{3}$. 
Hence, the first term in \cref{eq:differential:equation:a:tilde} scales linearly with $\tauzero$ and we cannot neglect $\ddot{\tildea}$.
Nevertheless, this flawed approach still contains the essence of the Landau-Zener effect. 

Indeed, we find from \cref{eq:differential:equation:a:tilde:limit} the expression
\begin{align}
    \label{eq:ansatz:a:ln:without:s}
    \tildea_a(\tau) = \exp\left(\frac{\ii}{2\epsilon}\ln \tau\right)
\end{align}
which displays a logarithmic phase and satisfies the initial condition \cref{eq:initial:condition:scaled:a}.

For positive values of $\tau$, that is for negative times of $t$ the logarithm is well-defined.
However, for negative values of $\tau$, that is for positive times of $t$, the multivaluedness of the logarithm becomes important, and a branch cut has to be introduced to make the logarithm unique. 

In this way, the Landau-Zener result is determined by $\tildea_a(-1)$ and with the choice $\ln\left(-1\right) = \ii\pi$, we arrive at the expression 
\begin{align}
    \tildea_a(-1) = a_{LZ}.
\end{align}

We emphasize that the choices $\ln\left(-1\right) = (2k+1)\ii\pi$ with $k\in\mathbb{Z}\setminus\{0\}$ do not lead to the Landau-Zener expression. 
Next, we show that the differential equation,  \cref{eq:differential:equation:a:tilde}, determine the correct value of $\ln(-1)$.

\section{Path in the complex plane}
\label{section:3}

In the preceding section, we have presented a heuristic argument in favor of a logarithmic phase singularity determining the Landau-Zener result, \cref{intro:eq:LZ}. 
We now confirm the appearance of the logarithm by discussing the time-dependence of the probability amplitude $\tildea$ given by \cref{eq:differential:equation:a:tilde} in terms of a path in the complex plane with a simple pole at the origin. 
The asymptotic behavior of this path brings to light the logarithm and promotes in this way the heuristic argument to a more rigorous one.

\subsection{The path and its asymptotics}

In this section, we first derive a differential equation for the path and solve it numerically. 
Next, we present approximate analytical solutions in the two time domains of positive and negative values of $\tau$, and discuss the limit $\tauzero\rightarrow\infty$.

\subsubsection{Probability amplitude represented by a path}

We start by generalizing, \cref{eq:ansatz:a:ln:without:s}, and make the ansatz
\begin{align}
    \label{eq:ansatz:a:ln}
    \tilde{a}(\tau; \tauzero) \equiv \exp\left[\frac{\ii}{2\epsilon}\int_{\mathcal{S}}\dd s\frac{1}{s}\right] = \exp\left[\frac{\ii}{2\epsilon}\ln\mathcal{S}\right]
\end{align}
where the path $\mathcal{S}$ in the complex plane is defined in such a way that the right-hand side of \cref{eq:ansatz:a:ln} is identical to the solution of the differential equation, \cref{eq:differential:equation:a:tilde}, subjected to the initial conditions \cref{eq:initial:conditon:tilde:a:scaled,eq:initial:condition:dot:tilde:a}.

Due to the ansatz, \cref{eq:ansatz:a:ln}, and the identity 
\begin{align}
    \label{eq:a:dot:identity}
    \dot{\tildea} = \frac{\ii}{2\epsilon}\frac{\dot{\mathcal{S}}}{\mathcal{S}}\tildea,
\end{align}
they translate into
\begin{align}
    \mathcal{S}(1; \tauzero) = 1
\end{align}
and 
\begin{align}
    \dot{\mathcal{S}}(1; \tauzero) = 0.
\end{align}

Hence, $\mathcal{S}$ connects the points \mbox{$\mathcal{S}(1; \tauzero)= 1$} and $\mathcal{S}(\tau; \tauzero)$ by a path determined by a differential equation following from \cref{eq:differential:equation:a:tilde} combined with the identity, \cref{eq:a:dot:identity}, and
\begin{align}
    \label{eq:connection:between:b:a}
    \ddot{\tildea} = \left\{\frac{\ii}{2\epsilon}\left[\frac{\ddot{\mathcal{S}}}{\mathcal{S}}-\left(\frac{\mathcal{\dot{S}}}{\mathcal{S}}\right)^{2}\right] -\frac{1}{4\epsilon^{2}}\left(\frac{\dot{\mathcal{S}}}{\mathcal{S}}\right)^{2}\right\}\tildea.
\end{align}

Indeed, when we substitute \cref{eq:a:dot:identity,eq:connection:between:b:a} into \cref{eq:differential:equation:a:tilde}, we arrive at the nonlinear differential equation
\begin{align}
    \label{eq:differential:equation:s}
    \frac{1}{\tauzero^{2}}\left[\frac{\ii}{2\epsilon}\frac{\ddot{\mathcal{S}}}{\mathcal{S}} - \left(\frac{1}{4\epsilon^{2}}+\frac{\ii}{2\epsilon}\right)\left(\frac{\dot{\mathcal{S}}}{\mathcal{S}}\right)^{2}\right] - \tau\frac{\dot{\mathcal{S}}}{\mathcal{S}} + 1 = 0  
\end{align}
of second order for the path $\mathcal{S}$.

\begin{figure*}[htbp]
    \includegraphics[width=\textwidth]{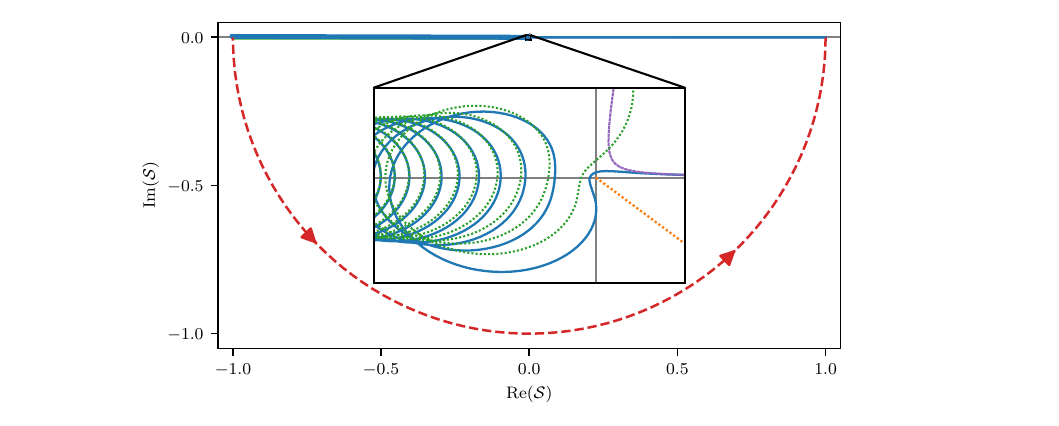}
    \centering
    \caption[]{Determination of the Landau-Zener transition probability amplitude $a$ from a path $\mathcal{S}$ in the complex plane with a pole at the origin obtained by numerically solving \cref{eq:differential:equation:s}. The exact path $\mathcal{S}$ starts at $\tau = 1$ from $\mathcal{S}(1; \tauzero) = 1$ and terminates at $\tau = -1$ at $\mathcal{S}(-1; \tauzero)$ avoiding the origin of the complex plane by moving to the upper half-plane as indicated by the inset.
    In the limit of $\tauzero\rightarrow\infty$ this path reduces to a straight line connecting the points $+1$ and $-1$ with a little detour to the imaginary axis in the neighborhood of the origin. When we close this asymptotic path by a half-circle of radius unity (red dashed line) the pole at the origin yields with the Cauchy integral theorem the value $\ii\pi$ for the path from $+1$ to $-1$. The avoidance of the origin already emerges from the approximate analytical expression, \cref{eq:S:approximation:neg:times}, for the path depicted by the dotted purple line in the inset.
    The green dotted line shows the approximate analytical expression, \cref{eq:S:large:positive:times}, and the orange dotted line represents the branch cut of the logarithm appropriate for the present situation. 
    Here, we used the parameters $\tauzero = 250.0$ and $\epsilon = 4.0$.}
    \label{fig:1}
\end{figure*}

In \cref{fig:1}, we show $\mathcal{S}$ for a fixed value of $\tauzero$ by numerically integrating \cref{eq:differential:equation:s}.
We note that $\mathcal{S}$ starts at $\tau = 1$ from $\mathcal{S} = 1$, and for decreasing positive values of $\tau$, that is for decreasing negative times $t$, avoids the origin of the complex plane by going into the upper half-plane.

For negative values of $\tau$, that is positive times $t$, the path returns to the real axis performing oscillations around it with a constant amplitude.
These drifting rotations in the contour $\mathcal{S}$ in the complex plane are the manifestation of the Stueckelberg oscillations \cite{Glasbrenner2025} in the probability amplitude $\tildea$.

\subsubsection{Avoidance of the pole and normalization condition}

We now present approximate analytical expressions for the path $\mathcal{S}$ for large values of $\tauzero$ in two different domains of $\tau$.
These formulae allow us in the next section to determine the asymptotic path $\mathcal{S}_a$. We start the discussion by considering the case $\tau\rightarrow 1$.

By direct differentiation, we can verify that
\begin{align}
    \label{eq:S:approximation:neg:times}
    \mathcal{S}(\abs{\tau}; \tauzero) \cong \abs{\tau}\left[1 + \frac{1}{\tauzero^{2}}g^{(+)}(\tau)\right]
\end{align}
is a solution of \cref{eq:differential:equation:s} including terms of the order $1/\tauzero^{2}$. 
Here, we have introduced the function 
\begin{align}
    g^{(+)}(\tau) \equiv \left(\frac{1}{8\epsilon^{2}} + \frac{\ii}{4\epsilon}\right)\frac{1}{\tau^{2}}\left(1-\tau^{2}\right)
\end{align}
which depends on $\tau$ but is independent of $\tauzero$.

In the inset of \cref{fig:1}, we compare and contrast the exact numerical expression for $\mathcal{S}$, depicted by the blue solid line and the approximation, \cref{eq:S:approximation:neg:times}, shown by purple dots, and find excellent agreement for positive values of $\tau$, that is large negative times $t$. 
In the neighborhood of $\tau=0$, $g^{(+)}$ diverges and thus the approximate path approaches a line under the angle $\arctan(2\epsilon)$ with respect to the real axis, in contrast to the exact path which stays close to the real axis. 

Moreover, the approximate path runs in the upper half-plane due to the positive imaginary part in $g^{(+)}$.
This feature is a consequence of the probability interpretation of $\abs{a}^{2}$.

In order to bring out this connection most clearly, we start from \cref{eq:ansatz:a:ln} in the form
\begin{align}
    \tildea = \exp\left(\frac{\ii}{2\epsilon}\ln\mathcal{S}\right)
\end{align}
and decompose $\mathcal{S} \equiv\abs{\mathcal{S}}\exp\left(\ii\theta\right)$
into its amplitude $\abs{\mathcal{S}}$ and phase $\theta$. 
Here, we have chosen $\theta=0$ at $\tau = 1$, that is at the starting point of the path.

Indeed, in this representation 
\begin{align}
    \label{eq:decomposition:S}
    \tildea = \exp\left(-\frac{\theta}{2\epsilon}\right) \exp\left(\frac{\ii}{2\epsilon}\ln\abs{\mathcal{S}}\right)
\end{align}
the phase $\theta$ of $\mathcal{S}$ determines the amplitude $\abs{\tildea}$, and the amplitude $\abs{\mathcal{S}}$ determines the phase of $\tildea$.
Hence, only positive values of $\theta$ guarantee the normalization condition, $\abs{\tildea}^{2}\leq 1$.

\subsubsection{Stueckelberg oscillations in the path}

For positive values of $\tau$, the approximate expression, \cref{eq:S:approximation:neg:times}, for $\mathcal{S}$ follows immediately from the differential equation, \cref{eq:differential:equation:s}, by a straightforward perturbative expansion in powers of $1/\tauzero^{2}$. 
Unfortunately, such an approach does not work for negative values of $\tau$.

Indeed, in this domain the analog of $g^{(+)}$ depends on $\tauzero$, and the time derivatives $\dot{\mathcal{S}}$ as well as $\ddot{\mathcal{S}}$ lead to additional powers of $\tauzero$ complicating a perturbative ansatz. 
These properties emerge, when we obtain the path $\mathcal{S}$ from asymptotic expressions of the parabolic cylinder functions derived in Ref.~\cite{Glasbrenner2026}. 

According to appendix \ref{appendix:A}, we find for negative values of $\tau$, the approximate expression 
\begin{align}
\label{eq:S:large:positive:times}
    \mathcal{S}(-\abs{\tau}; \tauzero) = -\abs{\tau}\left[1+\frac{1}{\tauzero}g^{(-)}(-\abs{\tau}; \tauzero)\right] 
\end{align}
with the abbreviations 
\begin{widetext}
\begin{align}
    \begin{split}
        \label{eq:g:minus}
        g^{(-)}(-\abs{\tau}; \tauzero) \equiv \frac{1}{\ii}\sqrt{1-\ee^{-\frac{\pi}{\epsilon}}}\ee^{\frac{\pi}{2\epsilon}}\left[\frac{1}{\abs{\tau}}\ee^{-\ii\epsilon\tauzero^{2}\left(\tau^{2}-1\right)}\ee^{-\frac{\ii}{\epsilon}\ln\abs{\tau}}\ee^{\ii\varphi(\tauzero)} + \ee^{-\ii\varphi(\tauzero)}\right]
    \end{split}
\end{align}
\end{widetext}
and 
\begin{align}
    \begin{split}
        \label{eq:varphi}
        \varphi(\tauzero)\equiv \frac{\pi}{4}-\epsilon\tauzero^{2} 
        - \frac{1}{\epsilon}\ln\left(\sqrt{2\epsilon}\tauzero\right) + \arg\left[\Gamma\left(\frac{\ii}{2\epsilon}\right)\right].
    \end{split}
\end{align}
Here, we have neglected terms of the order $1/\tauzero^{2}$ and $\Gamma$ denotes the $\Gamma$--function. 

We note that in contrast to positive values of $\tau$, the path $\mathcal{S}$ for negative values of $\tau$ contains not only an explicit time dependence but also a dependence on $\tauzero$. 
In particular, the first term in the square bracket of $g^{(-)}$, \cref{eq:g:minus}, decays with $1/\abs{\tau}$ as $\tau\rightarrow 1$, and at the same time, creates a circular motion in the complex plane with an increasing frequency determined by $\tauzero^{2}$. 
These circles are the manifestation of the Stueckelberg oscillations in the path. 
The logarithmic phase is not of importance appropriately away from $\tau = 0$.

In \cref{fig:1}, we compare and contrast this approximate path defined by \cref{eq:S:large:positive:times} and shown by the green dotted line to the exact one indicated by the blue solid line.
We note that the approximation reflects the key feature of the exact curve. 
However, there are deviations which originate from the expansion in terms of $1/\tauzero$.

\subsubsection{Choice of the branch cut}

The representation, \cref{eq:decomposition:S}, of $\tildea$ in terms of the phase $\theta$ and the amplitude $\abs{\mathcal{S}}$ of $\mathcal{S}$, together with the probability interpretation of $\abs{\tildea}^{2}$ enforces the use of positive values of $\theta$, only. 
Moreover, \cref{fig:1} seems to suggest that the path $\mathcal{S}$ does not reach deep into the lower right quadrant, and in particular, circumvents the pole at the origin traveling through the remaining quadrants. 
Therefore, we can place the branch cut of the logarithm in the lower-right quadrant as indicated by the orange dotted line in the inset of \cref{fig:1}. 

However, we admit that we have not determined the precise value of the phase angle of the branch cut with respect to the positive real axis. 
For this purpose, we would need to understand all possible paths in their dependence of $\tauzero$.
However, we argue that we can place the branch cut slightly below the real positive axis. 

\begin{figure}[htbp]
    \includegraphics[width=\columnwidth]{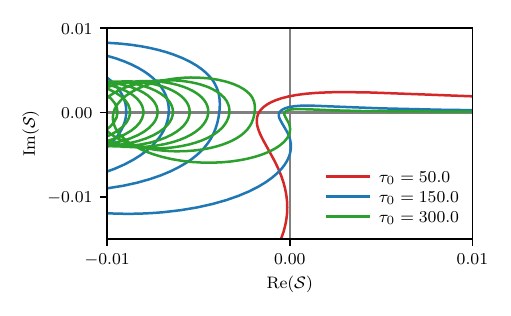}
    \centering
    \caption[]{Approach of the exact numerical path $\mathcal{S}$ to the asymptotic path $\mathcal{S}_a$ along the real axis connecting the points $+1$ and $-1$ in the complex plane for increasing values of $\tauzero$ in the neighborhood of the origin. 
    Here, we have depicted for a fixed value of $\epsilon = 4.0$ the cases: (i) $\tauzero = 50.0$ (red solid line), (ii) $\tauzero = 150.0$ (blue solid line) and (iii) (i) $\tauzero = 300.0$ (green solid line).}
    \label{fig:2}
\end{figure}

\subsection{The limit of the path}

We are now in the position to consider the influence of the parameter $\tauzero$ on the path $\mathcal{S}$. 
Again, our analysis relies on the exact numerical path as well as on the approximate analytic expressions, \cref{eq:S:approximation:neg:times,eq:S:large:positive:times}.

\subsubsection{Asymptotic path and a subtlety}

In \cref{fig:2}, we display the numerically obtained paths $\mathcal{S}$ for increasing values of $\tauzero$ and note that they move closer to the real axis.
Moreover, the excursions to the upper half-plane as well as of the amplitudes of the oscillations for negative values of $\tau$ decrease as well.

In the limit of $\tauzero\rightarrow\infty$, the resulting path connects the points $\mathcal{S} = 1$ and $\mathcal{S} = -1$ along the real axis with an infinitesimal detour to the positive imaginary axis. 

\begin{figure}[htbp]
    \includegraphics[width=\columnwidth]{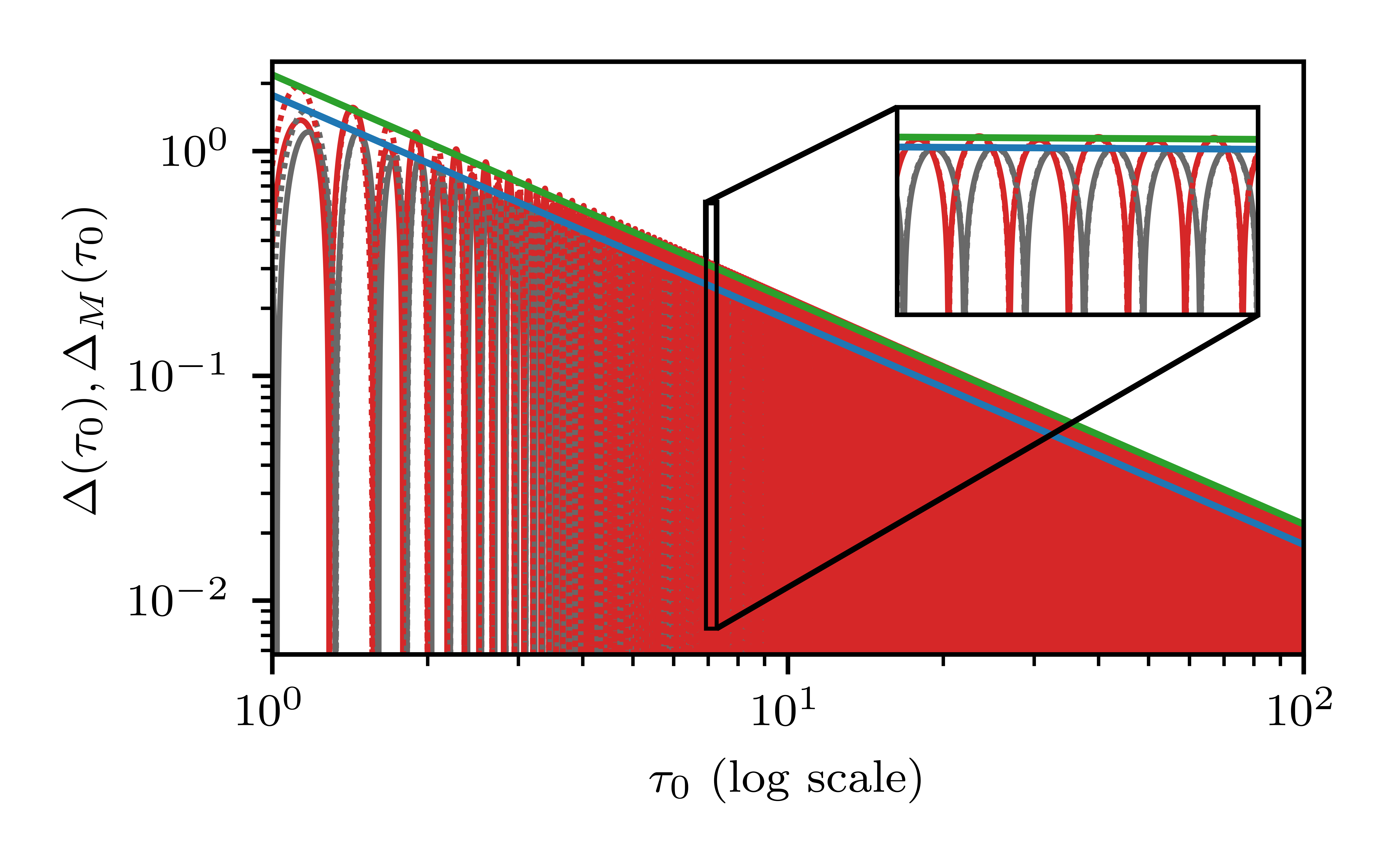}
    \centering
    \caption[]{Approach of the exact numerical points $\mathcal{S}(-1; \tauzero)$ and $\mathcal{S}_{M}(-1; \tauzero)$ in the complex plane towards $-1$ for increasing $\tauzero$ expressed by the distances $\Delta$ (red solid line) and $\Delta_M$ (gray solid line).
    Both curves follow a $1/\tauzero$-decay with oscillations around it as predicted by the approximate expressions, \cref{eq:Delta} and \cref{eq:Delta:M}, depicted by the red dotted and gray dotted lines. 
    Here, the solid green and blue tilted lines indicate the asymptotic amplitudes $\sqrt{\ee^{\pi/\epsilon}-1}$ and $\sqrt{\pi/\epsilon}$. 
    The phase shift of the oscillations survives even for large values of $\tauzero$ as indicated by the inset. 
    Here, we have used the value $\epsilon = 4.0$.}
    \label{fig:3}
\end{figure}

This behavior follows also immediately from \cref{eq:S:approximation:neg:times} and \cref{eq:S:large:positive:times} for $\tauzero\rightarrow\infty$. 
Although, we find for positive and negative values of $\tau$ a different approach towards the $\tauzero$-independent path given by $\tau$, we arrive at the asymptotic path
\begin{align}
    \label{eq:asymptotic:path:pos}
    \mathcal{S}_a(\tau) \equiv \limtauzeroinfty\mathcal{S}(\tau;\tauzero) = \tau,
\end{align}
for $\tau\neq 0$.
We emphasize that $\mathcal{S}_a$ does not go through the origin but avoids the pole by escaping into the upper half-plane. 

Hence, we obtain the differential equation 
\begin{align}
    \label{eq:approximated:ode:large:tauzero:new}
    \tau\frac{\dot{\mathcal{S}}_a}{\mathcal{S}_a} = 1
\end{align}
which is indeed the remnant of \cref{eq:differential:equation:s} independent of $\tauzero$. 

Indeed, we note from the definition, \cref{eq:g:minus}, of $g^{(-)}$ the oscillatory term, $\exp[-\ii\epsilon\tauzero^{2}(\tau^{2}-1)]$, which is responsible for the Stueckelberg oscillations. 
The differentiation of this term with respect to $\tau$ creates a contribution proportional to $\tauzero^{2}$. 
Hence, the correction term to $\dot{\mathcal{S}}$ given by $\dot{g}^{(-)}$ grows with $\tauzero$.

In contrast, when we first perform the limit $\tauzero\rightarrow\infty$, the correction term $g^{(-)}$ to $\mathcal{S}$ disappears, and a subsequent differentiation with respect to time only affects the term linear in $\tau$ which is independent of $\tauzero$. 

\subsubsection{Approach of the endpoint of the path}

Moreover, we emphasize that the approach of the final point $\mathcal{S}(-1; \tauzero)$ of the path $\mathcal{S}$ towards $-1$ is non-uniform. 
Indeed, for large but finite values of $\tauzero$, $\mathcal{S}(-1; \tauzero)$ is close to $-1$ but not identical to it, as shown by \cref{fig:3}.

Here, we display the distance
\begin{align}
    \label{eq:Delta:tauzero}
    \Delta(\tauzero) \equiv \abs{\mathcal{S}(-1; \tauzero) + 1}
\end{align}
of $\mathcal{S}(-1; \tauzero)$ from $-1$ for increasing $\tauzero$. 

We note a decay with $1/\tauzero$ and oscillations around it, in agreement with \cref{eq:S:large:positive:times} which when substituted into \cref{eq:Delta:tauzero} yields 
\begin{align}
    \label{eq:Delta}
    \Delta(\tauzero) = \frac{2}{\tauzero}\sqrt{1-\ee^{-\frac{\pi}{\epsilon}}}\ee^{\frac{\pi}{2\epsilon}}\abs{\cos\left[\varphi(\tauzero)\right]}.
\end{align}

Only in the limit of $\tauzero\rightarrow\infty$, do we find $\mathcal{S}(-1; \tauzero) = -1$ as predicted by \cref{eq:asymptotic:path:pos}. 
This property will become important in determining the Landau-Zener formula as discussed in the next section. 

\subsection{Cauchy theorem and Landau-Zener formula}

In the preceding section, we have shown that in the limit $\tauzero\rightarrow\infty$ the end point of the path $\mathcal{S}$ is at $-1$. 
Since $\mathcal{S}$ starts at $+1$ and avoids the pole in the \textit{upper} half-plane, the closed contour integral obtained by adding a half-circle in the lower half-plane as indicated in \cref{fig:1}, yields with the help of the Cauchy integral theorem the identity
\begin{align}
    \label{eq:lim:cauchy}
    \lim_{\tauzero \rightarrow \infty} \int_{\mathcal{S}(-1;\tau_0)} \frac{1}{s}\, \dd s = \ii\pi.
\end{align}
With \cref{eq:ansatz:a:ln} this result leads us immediately to the Landau-Zener formula, \cref{eq:Lz:border}.

This elementary derivation brings out two crucial ingredients of the factor $\pi$ in the Landau-Zener formula: (i) The phase $+\pi$ is encoded in the differential equation, \cref{eq:differential:equation:s}, forcing the path $\mathcal{S}$ to avoid the origin by moving to the upper half-plane. 
(ii) Since $\mathcal{S}$ needs to terminate at the point $-1$ in order to obtain the value $\pi$, the Landau-Zener formula is only valid in the limit $\tauzero\rightarrow\infty$.

\section{Markov path}
\label{section:4}

Our approach based on the asymptotic scaling, \cref{eq:scaling:t:tzero}, also answers the question posed in Ref.~\cite{Glasbrenner2023}: Why does the Markov approximation \cite{scullyzubairy1997, Paulisch2014} yield  the exact Landau-Zener result?

\subsection{Markov path and its asymptotics}

In order to resolve this conundrum, we follow in the present section arguments similar to the ones employed in section \ref{section:3}.
We first provide an explicit expression for the Markov path $\mathcal{S}_M$ and then show that it displays the same qualitative features as the exact path $\mathcal{S}$. 
In particular, we demonstrate that it avoids the origin by moving into the upper half-plane, displays the Stueckelberg oscillations, and its endpoint approaches the point $-1$ for $\tauzero\rightarrow\infty$.

\subsubsection{Explicit expression}

In contrast to the exact path $\mathcal{S}$ which is defined by the differential equation, \cref{eq:differential:equation:s}, the Markov path $\mathcal{S}_M$ is given explicitly. 
Indeed, when we recall \cite{Glasbrenner2025} the Markov solution 
\begin{align}
    \label{eq:markov:solution}
    \tildea_{M}(\tau; \tauzero) \equiv \ee^{\ii\mathrm{H}_{M}(\tau; \tauzero)}
\end{align}
with
\begin{align}
    \label{eq:h:m}
    \mathrm{H}_{M}(\tau; \tauzero) \equiv \ii\tauzero^{2}\int_{1}^{\tau}\dd\tauprime\ee^{-\ii\epsilon\tauzero^{2}\tau^{\prime 2}}\int_{1}^{\tauprime}\dd\tau^{\prime\prime}\ee^{\ii\epsilon\tauzero^{2}\tau^{\prime\prime 2}},
\end{align}
a comparison with \cref{eq:ansatz:a:ln} immediately provides us with the explicit expression
\begin{align}
    \label{eq:markov:trajectory}
    \mathcal{S}_{M}(\tau; \tauzero) \equiv \ee^{2\epsilon \mathrm{H}_{M}(\tau; \tauzero)}
\end{align}
for the path $\mathcal{S}_M$.
\begin{figure*}[htbp]
    \includegraphics[width=\textwidth]{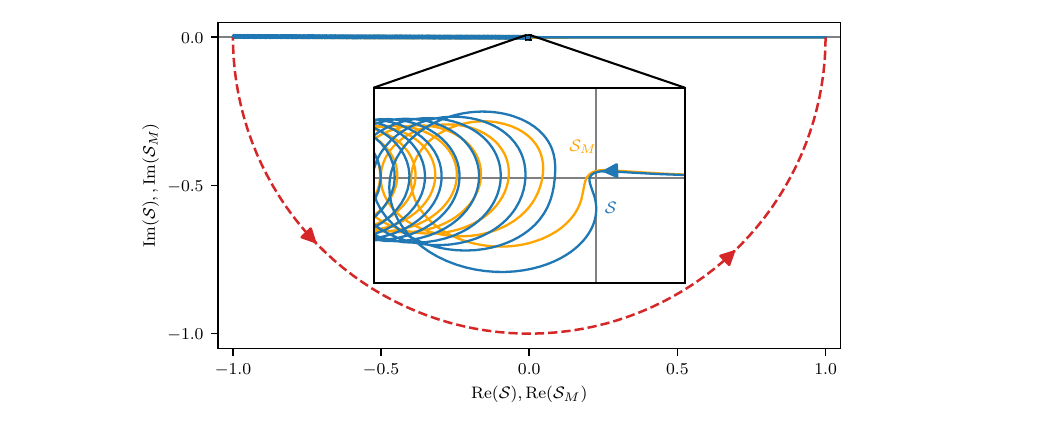}
    \centering
    \caption[]{Emergence of the exact Landau-Zener result, \cref{intro:eq:LZ} from the Markov approximation explained by the comparison between the Markov path $\mathcal{S}_M$ (solid orange line) given by \cref{eq:markov:trajectory} and the exact path $\mathcal{S} $ (solid blue line) following from \cref{eq:differential:equation:s}. Although both paths are different, they both avoid the pole at the origin by moving into the upper half-plane. Moreover, their corresponding asymptotic paths are identical as shown by \cref{eq:asymptotic:path:pos,eq:asymptotic:solution:s}. Hence, when we close the path in the lower half-plane by a half-circle we obtain with the help of the Cauchy integral theorem the contribution $\ii\pi$ for both paths. Here, we have chosen the values $\epsilon = 4.0$ and $\tauzero = 250.0$.}
    \label{fig:4}
\end{figure*} 

In \cref{fig:4}, we compare and contrast the Markov path $\mathcal{S}_M$ with the exact path $\mathcal{S}$. Both start at the point $+1$ and avoid the pole at the origin by moving into the upper-half plane. 
However, the two paths are slightly different. 

\subsubsection{Avoidance of the pole}

This escape of $\mathcal{S}_M$ also emerges analytically when we derive a differential equation for $\mathcal{S}_M$ analogous to \cref{eq:differential:equation:s}. 
Indeed, from Eqs.~(\ref{eq:h:m}) and (\ref{eq:markov:trajectory}), we find by differentiation the nonlinear differential equation of second order
\begin{align}
    \label{eq:differential:equation:s:m}
    \frac{1}{\tauzero^{2}}\left[\frac{\ii}{2\epsilon}\frac{\ddot{\mathcal{S}}_{M}}{\mathcal{S}_{M}} - \frac{\ii}{2\epsilon}\left(\frac{\dot{\mathcal{S}}_{M}}{\mathcal{S}_{M}}\right)^{2}\right] - \tau \frac{\dot{\mathcal{S}}_{M}}{\mathcal{S}_{M}} + 1 = 0.
\end{align}

When we compare this equation for the Markov path $\mathcal{S}_M$ to the one, \cref{eq:differential:equation:s} for the exact path $\mathcal{S}$, we note that apart from the contribution $1/(4\epsilon^{2})(\dot{\mathcal{S}}/\mathcal{S})^{2}$, they are identical.
In particular, the term $\ii/(2\epsilon)(\dot{\mathcal{S}}/\mathcal{S})^{2}$ is still present and positive, and ensures that $\mathcal{S}_M$ avoids the pole by escaping to the upper half-plane.

Indeed, when we neglect in \cref{eq:S:approximation:neg:times} the term $1/(8\epsilon^{2})$, we obtain the approximate Markov path
\begin{align}
    \label{eq:S:M:approximation:neg:times}
    \mathcal{S}_{M}(\abs{\tau}; \tauzero) \cong \abs{\tau}\left[1 + \frac{1}{\tauzero^{2}}g_{M}^{(+)}(\tau)\right]
\end{align}
with the positive imaginary part
\begin{align}
    g_{M}^{(+)}(\tau) \equiv \frac{\ii}{4\epsilon}\frac{1}{\tau^{2}}\left(1-\tau^{2}\right),
\end{align}
valid for positive values of $\tau$.

\subsubsection{Stueckelberg oscillations in the Markov path}

In complete analogy to the exact path $\mathcal{S}$, we derive in appendix \ref{appendix:A} the approximate expression 
\begin{align}
    \begin{split}
    \label{eq:S:M}
    \mathcal{S}_{M}(-\abs{\tau};\tauzero) = -\abs{\tau}\left[1 + \frac{1}{\tauzero}g_{M}^{(-)}(-\abs{\tau}; \tauzero)\right]
    \end{split}
\end{align}
for the Markov path valid for negative values of $\tau$. 
Here, we have introduced the abbreviations
\begin{align}
    \label{eq:g:minus:markov}
    g_{M}^{(-)}(-\abs{\tau}; \tauzero) \equiv \sqrt{\frac{\pi}{\epsilon}}\left[\frac{1}{\abs{\tau}}\ee^{\-\ii\epsilon\tauzero^{2}\left(\tau^{2} - 1\right)}\ee^{\ii\varphi_{M}(\tauzero)} - \ee^{-\ii\varphi_{M}(\tauzero)}\right]
\end{align}
and 
\begin{align}
    \label{eq:varphi:M}
    \varphi_{M}(\tauzero) \equiv \frac{\pi}{4}-\epsilon\tauzero^{2}.
\end{align}

Although the expressions for $g^{(-)}$ and $g_{M}^{(-)}$ given by \cref{eq:g:minus,eq:g:minus:markov} are similar, they differ in three characteristic features. The Markov approximation (i) replaces the amplitude  
$\sqrt{1-\ee^{-\pi/\epsilon}}\ee^{\pi/(2\epsilon)}$ of $g^{(-)}$ by $\sqrt{\pi/\epsilon}$;
(ii) retains only the first two terms of $\varphi$ as demonstrated by \cref{eq:varphi,eq:varphi:M}; and
(iii) changes the \textit{sum} of the terms of the square brackets in $g^{(-)}$ to the \textit{difference} of the corresponding terms in $g_{M}^{(-)}$.  

\subsubsection{Approach of the endpoint of the Markov path}

We are now in the position to analyze the approach of the Markov path $\mathcal{S}_M$ to $-1$ as defined by the distance 
\begin{align}
    \Delta_{M}(\tauzero) \equiv \abs{\mathcal{S}_{M}(-1; \tauzero) + 1}.
\end{align}

Indeed, when we substitute \cref{eq:S:M} into this expression we find
\begin{align}
    \label{eq:Delta:M}
    \Delta_M(\tauzero) = \frac{2}{\tauzero}\sqrt{\frac{\pi}{\epsilon}}\abs{\sin\left[\varphi_M(\tauzero)\right]}.
\end{align}

Hence, the distance $\Delta_M$ from $\mathcal{S}_M(-1;\tauzero)$ to $-1$ scales with $1/\tauzero$, in complete agreement with the asymptotic behavior of $\Delta$ given by \cref{eq:Delta}.

In \cref{fig:3}, we compare and contrast $\Delta_M$ and $\Delta$. 
Both share the scaling $1/\tauzero$ but the amplitudes of the oscillations are different as indicated by the two tilted parallel lines. 
Moreover, the inset shows the phase shift due to the different expressions for $\phi$ and $\phi_M$, and the substitution of the cosine by the sine function. 

\subsection{The limit of the Markov path}

The asymptotic limit $\tauzero\rightarrow\infty$ reduces \cref{eq:S:M:approximation:neg:times,eq:S:M} to the asymptotic path
\begin{align}
    \label{eq:asymptotic:solution:s}
    \mathcal{S}_{M,a}(\tau)\equiv\lim_{\tauzero\rightarrow\infty}\mathcal{S}_M(\tau; \tauzero) = \tau
\end{align} 
which is identical to the asymptotic exact path $\mathcal{S}_a$ given by \cref{eq:asymptotic:path:pos}.  
Hence, we arrive with the help the Cauchy theorem again at the result $a_{LZ}$ as shown in \cref{fig:4}.

In the application of the Cauchy theorem, it is again crucial that the point of termination of the Markov path is also $-1$, as predicted by both, \cref{eq:asymptotic:solution:s} and the limit 
\begin{align}
    \limtauzeroinfty \Delta_{M}(\tauzero) = 0
\end{align}
following from \cref{eq:Delta:M}. 

It is illuminating to rederive the endpoint $-1$ of the Markov path from the definition, \cref{eq:markov:trajectory}, which yields
\begin{align}
    \mathcal{S}_{M,a}(-1) = \limtauzeroinfty \exp\left[2\epsilon \mathrm{H}_{M}(-1;\tauzero)\right].
\end{align}

With the integral relation \cite{Glasbrenner2023, Glasbrenner2025}
\begin{align}
    \label{eq:double:integral}
    \int_{-\infty}^{\infty}\dd\tau\,\ee^{-\ii\epsilon\tau^{2}}\int_{-\infty}^{\tau}\dd\tau^{\prime}\ee^{\ii\epsilon\tau^{\prime 2}} = \frac{\pi}{2\epsilon},
\end{align}
we arrive at the result
\begin{align}
    \mathcal{S}_{M,a}(-1) = \ee^{\ii\pi} = -1.
\end{align}

In this way, we have provided an additional argument for the choice of the branch cut discussed in section \ref{section:2}.
Indeed, it is the double integral of \cref{eq:double:integral} together with the definition, \cref{eq:h:m}, which determines the phase $\ii\pi$, and therefore the Landau-Zener result, \cref{intro:eq:LZ}. 

\section{Conclusion and Outlook}
\label{section:5}

In this article, we have revisited the elementary Landau-Zener problem and have concentrated on the probability amplitude to remain in the initial state. 
Three key ideas define our approach: (i) The asymptotic scaling of the underlying equations,
(ii) the mapping of the dynamics onto a contour integral in the complex plane with a simple pole at the origin, and 
(iii) the asymptotic limit of this path.

The corresponding asymptotic path connects the points $+1$ and $-1$ and avoids the pole by moving into the upper half-plane.
When we close the path by a half-circle in the lower half-plane, the Cauchy theorem immediately yields with the logarithmic phase in the Landau-Zener transition probability amplitude the contribution $+\ii\pi$ and thus the Landau-Zener result.
Our analysis also emphasizes the importance of the asymptotic limit which allows a clear separation of the contributions to the Cauchy integral resulting from the path and the half-circle. 

Moreover, we have shown that the path defining the Markov solution displays an identical asymptotic behavior as the exact path. 
It is for this reason, that the Markov approximation reproduces the exact Landau-Zener result. 

Many extensions of our approach offer themselves. 
Here, we list only a few, and at the same time provide some guidance on possible solutions.

The exact path is defined by a nonlinear differential equation of second order which on first sight seems to simplify considerably in the asymptotic limit. 
However, the Stueckelberg oscillations ruins this naive strategy since the derivatives involve increasing powers of $\tauzero$. 

We are convinced that this problem would disappear if the path would be defined by an \textit{integral} rather then a \textit{differential} equation since the oscillations would average out. 
The integral equation for the probability amplitude following from the original set of differential equations might serve as an excellent starting point for the search for such an integral equation. 

Our analysis has shown that the Landau-Zener formula is a consequence of a logarithmic phase singularity and is determined by $\ln(-1)$. 
This observation is in agreement with Ref.~\cite{rojo2010} where the nested integrals of oscillatory functions, representing the probability amplitude in a perturbation series starting from a time-ordered product, have been performed in an exact way and resumed. 
The resulting exponential involves a contour integral with a path sightly above the real axis and a pole at the origin of the complex plane. 

The nested integrals extend from $-\infty$ to $+\infty$.
When we introduce our asymptotic scaling, we suspect that the integrals, now containing \textit{rapidly} oscillating functions, can be performed to rederive the result of Ref.~\cite{rojo2010}.
We claim that this derivation will again bring out most clearly the origin of the Landau-Zener effect.

Throughout our article, we have concentrated exclusively on the probability amplitude $a$. 
Thus, the natural question arises: Is our asymptotic scaling technique able to obtain an expression for $b$? 

On first sight we argue that the answer is 'no' since the asymptotic value of $b$ is intimately connected to the crossing of a Stokes line \cite{Zhu1992, Kayanuma1997, nakamura2012}.
However, one observation might offer some hope for a positive answer: The phase of the oscillations of the approach of $\Delta$ towards $-1$ is identical to the asymptotic phase of $b$. 

We postpone the complete answers to this and the other questions to a future publication.

\begin{acknowledgments}
We thank S. M. Barnett, M. A. Efremov, B. Garraway, S. Katolla and S. Varro for many fruitful discussions.
W.P.S. is most grateful to Texas A\&M University for a Faculty Fellowship at the Hagler Institute for Advanced Study at Texas \mbox{A\& M} University and to Texas \mbox{A\&M} AgriLife for the support of this work.
\end{acknowledgments}

\appendix
\onecolumngrid

\section{Asymptotic paths}
\label{appendix:A}
In this appendix, we derive expressions for the asymptotic paths $\mathcal{S}$ and $\mathcal{S}_M$, where we retain terms proportional to $1/\tauzero$.
Moreover, we focus on negative values of $\tau\rightarrow -1$ and build on results derived in Ref. \cite{Glasbrenner2025, Glasbrenner2026}.
Since $\tau$ in the present article is different from the one in Ref. \cite{Glasbrenner2025, Glasbrenner2026}, we have to first reexpress the formulae. 

\subsection{Asymptotics of exact solution}

In the scaling of the present article, the expression for the probability amplitude $a$ in the limit of negative values of $\tau$, that is for positive times $t$ reads
\begin{align}
    \label{eq:appendix:prob:a:in:terms:of:f}
    a(-\abs{\tau}; \tauzero) = \frac{f(-\abs{\tau}; \tauzero)}{f(1;\tauzero)}\left[1 + \frac{\ii}{2\epsilon}\frac{1}{\tauzero}g^{(-)}(-\abs{\tau}; \tauzero)\right]
\end{align}
where we have introduced the elementary wave 
\begin{align}
    \label{eq:appendix:elementary:wave:f}
    f(\tau; \tauzero) \equiv \ee^{\ii\epsilon\tauzero^{2}\tau^{2}/2}\ee^{\frac{\ii}{2\epsilon}\ln\left(\tau\tauzero\right)}
\end{align}
and the abbreviation
\begin{align}
    \label{eq:appendix:g:minus}
    g^{(-)}(-\abs{\tau}; \tauzero) \equiv \frac{1}{\ii}\sqrt{1-\ee^{-\frac{\pi}{\epsilon}}}\left[\frac{1}{\abs{\tau}}\ee^{-\frac{\pi}{2\epsilon}}\ee^{\ii\phi}\frac{f^{\ast}(-\abs{\tau}; \tauzero)}{f(-\abs{\tau};\tauzero)} + \ee^{\frac{\pi}{2\epsilon}}\ee^{-\ii\phi}\frac{f(1; \tauzero)}{f^{\ast}(1; \tauzero)}\right]
\end{align}
with the definition
\begin{align}
    \label{eq:appendix:phi}
    \phi \equiv \frac{\pi}{4} - \frac{1}{2\epsilon}\ln\left(2\epsilon\right) + \arg\left[\Gamma\left(\frac{\ii}{2\epsilon}\right)\right]
\end{align}
and $\Gamma$ denotes the $\Gamma$-function.

When we substitute the definition, \cref{eq:appendix:elementary:wave:f}, of $f$ into the expression, 
\cref{eq:appendix:prob:a:in:terms:of:f}, for $a$, bring the square brackets in the exponent and combine it with the logarithm in $f$ using the relation $\ln\left(x\right)+\beta \approx \ln\left[x(1+\beta)\right]$, we arrive at the formula 
\begin{align}
    a(-\abs{\tau}; \tauzero) \cong \ee^{\ii\epsilon\tauzero^{2}\left(\tau^{2}-1\right)/2}\ee^{\frac{\ii}{2\epsilon}\ln\left[\mathcal{S}(-\abs{\tau};\tauzero)\right]}
\end{align}
where 
\begin{align}
    \mathcal{S}(-\abs{\tau}; \tauzero) = -\abs{\tau}\left[1+\frac{1}{\tauzero}g^{(-)}(-\abs{\tau}; \tauzero)\right]. 
\end{align}

We conclude by substituting the expression, \cref{eq:appendix:elementary:wave:f}, for $f$ into the definition, \cref{eq:appendix:g:minus}, of $g^{(-)}$ and obtain the formula
\begin{align}
    g^{(-)}(-\abs{\tau}; \tauzero) = \frac{1}{\ii}\sqrt{1-\ee^{-\frac{\pi}{\epsilon}}}\ee^{\frac{\pi}{2\epsilon}}\left[\frac{1}{\abs{\tau}}\ee^{-\ii\epsilon\tauzero^{2}\tau^{2}}\ee^{-\frac{\ii}{\epsilon}\ln\left(\abs{\tau}\tauzero\right)}\ee^{\ii\phi}+ \ee^{\ii\epsilon\tauzero^{2}}\ee^{\frac{\ii}{\epsilon}\ln\left(\tauzero\right)}\ee^{-\ii\phi}\right].
\end{align}

For $\tau = -1$ this expression reduces to 
\begin{align}
    \label{eq:appendix:g:minus:limit}
    g^{(-)}(-1; \tauzero) = \frac{2}{\ii}\sqrt{1-\ee^{-\frac{\pi}{\epsilon}}}\ee^{\frac{\pi}{2\epsilon}}\cos\left[\varphi(\tauzero)\right]
\end{align}
with
\begin{align}
    \varphi(\tauzero) \equiv \frac{\pi}{4} - \epsilon\tauzero^{2} - \frac{1}{\epsilon}\ln\left(\sqrt{2\epsilon}\tauzero\right) + \arg\left[\Gamma\left(\frac{\ii}{2\epsilon}\right)\right]
\end{align}
where we have recalled the definition, \cref{eq:appendix:phi}, of $\phi$.

\subsection{Markov approximation}
In the scaling of the present article, the Markov solution \cite{Glasbrenner2025} reads
\begin{align}
    a_{M}(\tau; \tauzero)\equiv\ee^{\ii\epsilon\tauzero^{2}\left(\tau^{2}-1\right)/2}\ee^{\ii \mathrm{H}_{M}(\tau; \tauzero)}
\end{align}
with
\begin{align}
    \mathrm{H}_{M}(\tau; \tauzero) \equiv \ii\tauzero^{2}\int_{1}^{\tau}\dd\tauprime\ee^{-\ii\epsilon\tauzero^{2}\tau^{\prime 2}}\int_{1}^{\tauprime}\dd\tau^{\prime\prime}\ee^{\ii\epsilon\tauzero^{2}\tau^{\prime\prime 2}},
\end{align}
and leads us to the expression
\begin{align}
    \label{eq:appendix:markov:trajectory}
    \mathcal{S}_{M}(\tau; \tauzero) = \ee^{2\epsilon \mathrm{H}_{M}(\tau; \tauzero)}
\end{align}
for the Markov path. 

We recall from Ref.~\cite{Glasbrenner2025} the identity
\begin{align}
    \label{appendixB:eq:connection:formula:H:M}
    \mathrm{H}_{M}(-\abs{\tau}; \tauzero) = \mathrm{H}_{M}(\abs{\tau}; \tauzero) + \ii\tauzero^{2}\mathcal{F}(-1; \tauzero)\int_{0}^{-\abs{\tau}}\dd\tauprime\ee^{-\ii\epsilon\tauzero^{2}\tau^{\prime 2}}
\end{align}
with the definition 
\begin{align}
    \mathcal{F}(\tau; \tauzero) \equiv \int_{1}^{\tau}\dd\tauprime\ee^{\ii\epsilon\tauzero^{2}\tau^{\prime 2}}.
\end{align}

In the asymptotic limit of $\abs{\tau}\rightarrow 1$ the asymptotic formula \cite{Glasbrenner2025}
\begin{align}
    \mathrm{H}_{M}(\abs{\tau}; \tauzero) \cong \frac{1}{2\epsilon}\ln\left(\abs{\tau}\right) 
\end{align}
for $\mathrm{H}_M$ is independent of $\tauzero$, and when we insert \cref{appendixB:eq:connection:formula:H:M} into \cref{eq:appendix:markov:trajectory}, we arrive at
\begin{align}
    \begin{split}
    \label{appendixB:eq:S:M}
    \mathcal{S}_{M}(-\abs{\tau};\tauzero) \cong \abs{\tau}\exp\left(2\ii\epsilon\tauzero^{2}\mathcal{F}(-1;\tauzero)\int_{0}^{-\abs{\tau}}\dd\tauprime\ee^{-\ii\epsilon\tauzero^{2}\tau^{\prime 2}}\right).
    \end{split}
\end{align}

For $\tau\rightarrow-1$, we find the asymptotic expansions
\begin{align}
    \mathcal{F}(-1; \tauzero) = \int_{1}^{-1}\dd\tauprime\ee^{\ii\epsilon\tauzero^{2}\tau^{\prime 2}}&\cong -\frac{1}{\tauzero}\left[ \sqrt{\frac{\ii\pi}{\epsilon}} + \frac{\ee^{\ii\epsilon\tauzero^{2}}}{\ii\epsilon\tauzero}\right]
\end{align}
and
\begin{align}
    \int_{0}^{-\abs{\tau}}\dd\tauprime\ee^{-\ii\epsilon\tauzero^{2}\tau^{\prime 2}} \cong -\frac{1}{2\tauzero}\left[\sqrt{\frac{(-\ii)\pi}{\epsilon}} - \frac{\ee^{-\ii\epsilon\tauzero^{2}\tau^{2}}}{\ii\epsilon\tauzero\abs{\tau}}\right],
\end{align}
which reduce \cref{appendixB:eq:S:M} to the expression
\begin{align}
    \begin{split}
    \label{appendixB:eq:S:M:limit}
    \mathcal{S}_{M}(-\abs{\tau};\tauzero) \cong -\abs{\tau}\left[1 + \frac{1}{\tauzero}g_{M}^{(-)}(-\abs{\tau}; \tauzero)\right],
    \end{split}
\end{align}
with 
\begin{align}
    g_{M}^{(-)}(-\abs{\tau}; \tauzero) \equiv \sqrt{\frac{\pi}{\epsilon}}\left[\frac{1}{\abs{\tau}}\ee^{\ii(\pi/4-\epsilon\tauzero^{2}\tau^{2})} - \ee^{-\ii(\pi/4 -\epsilon\tauzero^{2})}\right].
\end{align}

For $\tau = -1$, we find
\begin{align}
    g_M^{(-)}(-1;\tauzero) = -\frac{2}{\ii}\sqrt{\frac{\pi}{\epsilon}}\sin\left[\varphi_M(\tauzero)\right]
\end{align}
with
\begin{align}
    \varphi_M(\tauzero) \equiv \frac{\pi}{4} -\epsilon\tauzero^{2} .
\end{align}

\nocite{*}

\twocolumngrid

\bibliography{references}

\end{document}